# Decrease of Entropy and Chemical Reactions

*Yi-Fang Chang*

*Department of Physics, Yunnan University, Kunming, 650091, China*

(e-mail: yifangchang1030@hotmail.com)

**Abstract**

The chemical reactions are very complex, and include oscillation, condensation, catalyst and self-organization, etc. In these case changes of entropy may increase or decrease. The second law of thermodynamics is based on an isolated system and statistical independence. If fluctuations magnified due to internal interactions exist in the system, entropy will decrease possibly. In chemical reactions there are various internal interactions, so that some ordering processes with decrease of entropy are possible on an isolated system. For example, a simplifying Fokker-Planck equation is solved, and the hysteresis as limit cycle is discussed.

Key words: entropy, chemical reaction, internal interaction, oscillation, catalyst.
PACS: 05.70.-a, 82.60.-s, 05.20.-y.

**1.Introduction**

The measure of disorder is used by thermodynamic entropy. According to the second law of thermodynamics, the entropy of the universe seems be constantly increasing. A mixture of two pure substances or dissolve one substance in another is usually an increase of entropy. In a chemical reaction, when we increase temperature of any substance, molecular motion increase and so does entropy. Conversely, if the temperature of a substance is lowered, molecular motion decrease, and entropy should decreases. In nature, the general tendency is toward disorder.

Usual development of the second law of the thermodynamics was based on an open system, for example, the dissipative structure theory. Nettleton discussed an extended thermodynamics, which introduces the dissipative fluxes of classical nonequilibrium thermodynamics, and modified originally rational thermodynamics with nonclassical entropy flux [1]. The Gibbs equation from maximum entropy is a statistical basis for differing forms of extended thermodynamics.

Recently, Gaveau, et al., [2] discussed the variational nonequilibrium thermodynamics of reaction-diffusion systems. Cangialosi, et al. [3], provided a new approach to describe the component segmental dynamics of miscible polymer blends combining the configurational entropy and the self-concentration. The results show an excellent agreement between the prediction and the experimental data. Stier [4] calculated the entropy production as function of time in a closed system during reversible polymerization in nonideal systems, and found that the nature of the activity coefficient has an important effect on the curvature of the entropy production.

In particular, Cybulski, et al. [5], analyzed a system of two different types of Brownian particles confined in a cubic box with periodic boundary conditions. Particles of different types annihilate when they come into close contact. The annihilation rate is matched by the birth rate, thus the total number of each kind of particles is conserved. The system evolves towards those stationary distributions of particles that minimize the Renyi entropy production, which decreases monotonically during the evolution despite the fact that the topology and geometry of the interface exhibit abrupt and violent changes. Kinoshita, et al. [6], developed an efficient method to evaluate the translational and orientational contributions to the solute-water pair-correlation entropy that is a major component of the hydration entropy in an analytical manner. They studied the effects of the solute-water attractive interaction. A water molecule is modeled as a hard sphere of diameter, which becomes larger, the percentage of the orientational contribution first increases, takes a maximum value at depends on the strength of the solute-water attractive interaction, and then decreases toward a limiting value. The percentage of the orientational contribution reduces progressively as the solute-water attractive interaction becomes stronger. They discussed the physical origin of the maximum orientational restriction.

In fact, the second law of thermodynamics holds only for an isolated system, and it is a probability law. This shows that transition probability from the molecular chaotic motion to the regular motion of a macroscopic body is very small. The basis of thermodynamics is the statistics, in which a basic principle is statistical independence: The state of one subsystem does not affect the probabilities of various states of the other subsystems, because different subsystems may be regarded as weakly interacting [7]. It implies that various interactions among these subsystems should not be considered. But, if various internal complex mechanism and interactions cannot be neglected, perhaps a state with smaller entropy (for example, self-organized structure) will be able to appear. In these cases, the statistics and the second law of thermodynamics should be different [8-10]. For instance, the entropy of an isolated fluid whose evolution depends on viscous pressure and the internal friction does not increase monotonically [11].

According to the Boltzmann and Einstein fluctuation theories, all possible microscopic states of a system are equal-probable in thermodynamic equilibrium, and the entropy tends to a maximum value finally. In statistical mechanics fluctuations may occur, and always bring the entropy to decrease [7,12-14]. Under certain condition fluctuations can be magnified [12-14].

Chichigina [15] proposed a new method for describing selective excitation as the addition of information to a thermodynamic isolated system of atoms, decreasing the entropy of the system as a result. This information approach is used to calculate the light-induced drift velocity. The computational results are in good agreement with experimental data.

**2. Fokker-Planck Equation, Oscillation and Change of Entropy**

There are various internal interactions in chemical reactions, and here noise and internal fluctuations cannot be neglected often. The nonlinear chemical reactions belong to the non-equilibrium phase transformation, and fluctuations magnified may derive far-equilibrium states.

It is known that change of entropy may increase or decrease in chemical reactions [16,17]. If the chemical heterogeneous reaction, multistability and hysteresis loop become to an entropy function, which will be not a monotone increasing function, these phenomena will possess possibly new property of entropy.

Gaveau, et al., introduced a standard Master equation [2,18]:

$$\frac{\partial P(\{n_i\},t)}{\partial t} = \sum_r [W_r(\{n_i - r_i\})P(\{n_i - r_i\},t) - W_r(\{n_i\})P(\{n_i\},t)], \quad (1)$$

and an approximate Fokker-Planck equation:

$$\frac{\partial p(x,t)}{\partial t} = -\sum_{i=1}^{s} \frac{\partial}{\partial x_i}[A_i(x)p(x,t)] + \frac{1}{2V}\sum_{i,j=1}^{s} \frac{\partial^2}{\partial x_i \partial x_j}(D_{ij}p), \quad (2)$$

where $p(x,t)$ is the density of the probability distribution function. This is a kinetic equation of the distribution function [12,19].

We solve a simplifying Fokker-Planck equation:

$$\frac{dp}{dt} = -\alpha p - \beta p^3 + \gamma \Delta p . \tag{3}$$

1.When $\beta = \gamma = 0$, the distribution function is $p = Ce^{-\alpha t}$. It corresponds to the Maxwell-Boltzmann distribution, and its entropy is

$$s = -Kp \ln p = -KCe^{-\alpha t}(\ln C - \alpha t) . \tag{4}$$

Let $s_0 = -KC \ln C$, so

$$ds = s - s_0 = KC[\ln C(1 - e^{-\alpha t}) + \alpha t e^{-\alpha t}] . \tag{5}$$

Therefore, $ds > 0$ for $\alpha > 0$, and $ds < 0$ for $\alpha < 0$. 2.When $\gamma = 0$, $p^2 = \alpha/\beta(e^{2\alpha t + c} - 1)$. In this case the equation has two solutions: $p > 0$ and $p < 0$, but the latter is meaningless. It corresponds to the Bose-Einstein distribution [13,20], and its entropy is

$$s = K[(1+p)\ln(1+p) - p \ln p] > 0 . \tag{6}$$

Here $p = \sqrt{\alpha/\beta(e^{2\alpha t + c} - 1)}$ decreases with increasing t for $\alpha > 0$. When $p >> 0$, $s = K \ln p$. Let $s_0 = K[(1+p_0)\ln(1+p_0) - p_0 \ln p_0]$, where $p_0$ corresponds to $t_0 < t$, so

$$ds = s - s_0 = K \ln[\frac{(1+p)^{1+p}}{p^p} \frac{p_0^{p_0}}{(1+p_0)^{1+p_0}}] . \tag{7}$$

This is a decreasing function, $ds < 0$. It is consistent with the Bose-Einstein statistics corresponding condensation. Further, the general equations with the nonlinear term, which corresponds often to various interactions, may have different phase transformations.

In chemical reactions more intriguing is the Belousov-Zhabotinskii (BZ) class of oscillating reactions some of which can continue for hours [21-24]. Field and Dutt, et al., discussed the limit cycles and the discontinuous entropy production changes in the reversible Oregonator [25,26], which may describe the oscillations in the BZ reaction. The chemical oscillations as one type of the limit cycle imply at least that entropy cannot be increase monotonically.

The Fokker-Planck equation may be the reaction-diffusion equation. It can be extended to several equations. The two equations may describe Brusselator in chemical reactions. The three equations may describe Oregonator and BZ reaction. Dutt, et al., reported changes discontinuously at Hopf bifurcation points leading to limit cycle oscillations out of an unstable steady state [27].

Under some cases the factual changes of systems controlled by parametric values can be achieved through the existing dynamical process inside systems, in which hysteresis loop may be formed spontaneously. The character of the hysteresis loop can bring states of system to periodic change spontaneously with time.

Steady states in the Raleigh-Benard convection have exhibited hysteresis in entropy production [28]. At different steady states Dutt calculated entropy production [29], and the average entropy production for the unstable steady states generating limit cycle oscillations. The hysteresis in the entropy production as a function of the rate constant k of the steps in Oregonator (k decreasing or increasing) as time evolves showed the stable and unstable (limit cycle) phases. In the case the Floquet exponent is positive or negative for different critical points of Hopf bifurcations, i.e., the entropy productions have two values for the same chemical reaction. The entropy production is calculated in the model in a subcritical Hopf bifurcation region. Dutt, et al., calculated the entropy production for steady and oscillatory states in another region [27,30,31].

Dutt discussed the hysteresis in entropy production in a model exhibiting oscillations in the BZ reaction, and a bistability in the entropy production between a steady state and an oscillatory state in the neighborhood of a subcritical Hopf bifurcation. He identified this region in the past by a perturbation method. The numerical results present a thermodynamic formulation of multistable structures in a nonlinear dissipative system [29].

**3. Decrease of Entropy and Condensed Process**

We discussed fluctuations magnified due to internal interactions, etc., so that the equal-probability does not hold. In this case, the entropy with time would be defined as [10]

$$S(t) = -k \sum_r P_r(t) \ln P_r(t). \tag{8}$$

From this a possible decrease of entropy is calculated in an internal condensed process. Internal interactions, which bring about inapplicability of the statistical independence, cause possibly decreases of entropy in an isolated system. The possibility is researched for attractive process, internal energy, system entropy and nonlinear interactions, etc. An isolated system may form a self-organized structure [8-10].

For the systems with internal interactions, we proposed that the total entropy should be extended to [10]

$$dS = dS^a + dS^i, \tag{9}$$

where $dS^a$ is an additive part of entropy, and $dS^i$ is an interacting part of entropy. Eq.(9) is similar to a well known formula:

$$dS = d_i S + d_e S, \tag{10}$$

in the theory of dissipative structure proposed by Prigogine. Two formulae are applicable for internal or external interactions, respectively.

The entropy of non-ideal gases is:

$$S = S_{id} + N \log(1 - Nb/V). \tag{11}$$

This is smaller than one of ideal gases, since b is four times volume of atom, $b > 0$. This corresponds to the existence of interactions between the gas molecules, and average of forces between molecules is attractive [13]. The entropy of a solid is:

$$S = 2\pi^2 V T^3 / 15(\hbar \overline{u})^3, \tag{12}$$

so $dS < 0$ for $dT < 0$. The free energy with the correlation part of plasma is [13]:

$$F = F_{id} - \frac{2e^3}{3} \frac{\pi^{1/2}}{(VT)^{1/2}} \left( \sum_a N_a z_a^2 \right)^{3/2}. \tag{13}$$

Correspondingly, the entropy is:

$$dS = S - S_{id} = -\frac{e^3}{3} \frac{\pi^{1/2}}{V^{1/2}} \left( \sum_a N_a z_a^2 \right)^{3/2} T^{-3/2} < 0. \tag{14}$$

In these cases, there are the electric attractive forces between plasma.

According to

$$S = k \ln \Omega, \tag{15}$$

in an isolated system there are the n-particles, which are in different states of energy respectively, so $\Omega_1 = n!$. Assume that internal attractive interaction exists in the system, and the n-particles will cluster to m-particles. If they are in different states of energy still, there will be $\Omega_2 = m!$. Therefore, in this process

$$S_2 - S_1 = dS = k \ln(\Omega_2 / \Omega_1) = k \ln(m!/n!). \tag{16}$$

So long as $m < n$ for the condensed process, entropy decreases $dS < 0$. Conversely, $m > n$ for the dispersed process, entropy increases $dS > 0$. In these cases it is independent that each cluster includes energy more or less. In an isolated system, cluster number is lesser, the degree of irregularity and entropy are smaller also. This is consistent with a process in which entropy decreases from gaseous state to liquid and solid states.

In the chemical reactions the entropy changes may be calculated from the standard molar entropy [32]

$$S^0 = \sum v_p S^0(products) - \sum v_r S^0(reactions), \tag{17}$$

for example, for the conversion $2NO + O_2 \rightarrow 3NO_2, \Delta S^0 = -146.5 JK^{-1}$. Further, Petrucci, et al., proposed to be able to apply some qualitative reasoning: Because three moles of gaseous reactants produce only two moles of gaseous products, they can expect the entropy to decrease; that is that $\Delta S^0$ should be negative [17].

When gaseous ions change over to aquosity in hydration reactions, their standard molar entropy remarkably decreases. When ions are smaller or their charges increase, entropy of solution decreases more. This is consistent with our conclusion [8-10]. A dissolved process may be an automatic one.

Based on the thermodynamic relationships

$$\Delta G^0 = -RT \ln \beta, \tag{18}$$

and the change of the entropy

$$\Delta S^0 = (\Delta H^0 - \Delta G^0)/T, \tag{19}$$

the thermodynamic function $T\Delta S^0 = -19.91 kJmol^{-1}$ at $25^0 C$ for the complex $[Cd(NH_2Me)_4]^{2+}$ [33]. For $CH_4, CO_2, H_2O, O_3$, there is $\Delta H^0 < \Delta G^0$, i.e., $\Delta S^0 < 0$ [17]. These pointed out that change of entropy is negative value, namely, entropy decrease in these cases.

Moreover, the standard product entropy may be negative value in metal ions in aqueous solution [17,33].

The entropy is expressed in terms of the partition function Z as

$$S = kT\frac{\partial(\ln Z)}{\partial T} + k\ln Z. \tag{20}$$

For order and disorder the increase of entropy in fusion is

$$\Delta S_{fu} = kN = nR, \tag{21}$$

which is called the communal entropy [17]. Contrarily, the entropy in condensation should is

$$\Delta S_{con} = -kN = -nR, \tag{22}$$

which decreases. In the general case, fusion and condensation are in the open systems. But, a few condensations with internal interactions may be in some isolated systems. Further, the general phase transformation is in the open systems. But, a spontaneous magnetization may be in an isolated system. This is an ordering process, in which entropy should decrease.

**4.Catalyst and Discussion**

The catalysts may substantially lower the activation energy, and increase the rates of some chemical reactions without itself being consumed in the reactions, and may control direction of reactions. The mass of the catalyst remains constant [34,35].

In fact, the catalyst in chemical reactions possesses some character of the auto-control mechanism in an isolated system. If it does not need the input energy, at least in a given time interval, the self-catalyst is similar with auto-control like a type of Maxwell demon, which is just a type of internal interactions. The demon may be a permeable membrane. For the isolated system, this is possible that the catalyst and other substance are mixed to produce new order substance with smaller entropy. Ordering is the formation of structure through the self-organization from a disordered state.

Dutt proposed the result interesting to biophysical processes like glycolytic oscillations, and may be considered useful for better efficiency of the biochemical engines [27], namely if their structures is better, it will implies that there is smaller entropy for the same process.

In the chemical and biological self-organizing processes some isolated systems may tend to the order states spontaneously. Ashby pointed out [36]: Ammonia and hydrogen are two gases, but they mix to form a solid. There are about 20 types of amino acid in germ, they gather together, and possess a new reproductive property. Generally solid is more order than gas, and corresponding solid entropy is smaller than gaseous entropy. Germ should be more order than amino acid yet.

Prigogine and Stengers [14] discussed a case: When a circumstance of Dictyostelium discoideum becomes lack of nutrition, discoideum as some solitary cells will unite to form a big cluster spontaneously. In this case these cells and nutrition-liquid may be regarded as an isolated system.

Jantsch [37] pointed out: When different types of sponge and water are mixed up in a uniform suspension, they rest after few hours, and then separate different types automatically. It is a more interesting order process, a small hydra is cut into single cell, and these cells will spontaneously evolve, firstly form some cell-clusters, next form some malformations, finally they become a normal hydra.

The decrease of entropy relates possibly to the thermodynamics of enzyme-catalyzed reactions. An enzyme is a complex protein molecule that can act as a biological catalyst. In biology there are self-organize and self-assembly [38], etc.

These phenomena show that nature does not tend toward disorder always, and are some ordering processes with decrease of entropy.

Our conclusion is: The chemical reactions are complex processes. When there are various internal interactions and fluctuations in an isolated system: 1. It should be discussed that all middle change process from begin to end is always entropy increase. 2.The entropy increase principle in these processes may not hold always. 3. All of world does not trend to disorder always. Further, a negative temperature is based on the Kelvin scale and the condition $dU>0$ and $dS<0$. Conversely, there is also the negative temperature for $dU<0$ and $dS>0$. We find that the negative temperature derives necessarily decrease of entropy [39]. Moreover, the negative temperature is a query, and is contradiction with usual meaning of temperature and with some basic concepts of physics and mathematics. It is a notable question in nonequilibrium thermodynamics. In this paper, we attend mainly to thermodynamic results and change of entropy in the chemical reactions.